\documentclass{article}
\usepackage{booktabs}
\usepackage[T1]{fontenc} % add special characters (e.g., umlaute)
\usepackage[utf8]{inputenc} % set utf-8 as default input encoding
\usepackage{ismir,amsmath,cite,url}
\usepackage{graphicx}
\usepackage{color}
\usepackage{amssymb}
\usepackage{bm}
\usepackage{paralist}
\usepackage{lineno}
\title{Music2Latent: Consistency Autoencoders for Latent Audio Compression}

\multauthor
{Marco Pasini$^1$ \hspace{1.5cm} Stefan Lattner$^2$ \hspace{1.5cm} György Fazekas$^1$} {
$^1$Queen Mary University of London, UK \hspace{1cm}
$^2$Sony Computer Science Laboratories, Paris, France\\
{\tt m.pasini@qmul.ac.uk}
}

\def\authorname{M. Pasini, S. Lattner, and G. Fazekas}

\usepackage[bookmarks=false,pdfauthor={\authorname},pdfsubject={\papersubject},hidelinks]{hyperref}
\begin{document}

\maketitle

\begin{abstract}
Efficient audio representations in a compressed continuous latent space are critical for generative audio modeling and Music Information Retrieval (MIR) tasks. However, some existing audio autoencoders have limitations, such as multi-stage training procedures, slow iterative sampling, or low reconstruction quality.
We introduce Music2Latent, an audio autoencoder that overcomes these limitations by leveraging consistency models. Music2Latent encodes samples into a compressed continuous latent space in a single end-to-end training process while enabling high-fidelity single-step reconstruction. Key innovations include conditioning the consistency model on upsampled encoder outputs at all levels through cross connections, using frequency-wise self-attention to capture long-range frequency dependencies, and employing frequency-wise learned scaling to handle varying value distributions across frequencies at different noise levels.   
We demonstrate that Music2Latent outperforms existing continuous audio autoencoders in sound quality and reconstruction accuracy while achieving competitive performance on downstream MIR tasks using its latent representations. To our knowledge, this represents the first successful attempt at training an end-to-end consistency autoencoder model. Pretrained weights are available under [this link].\footnote{\href{https://github.com/SonyCSLParis/music2latent}{https://github.com/SonyCSLParis/music2latent}}
\end{abstract}

\section{Introduction}\label{sec:introduction}
The ability to faithfully and efficiently represent high-dimensional audio data in a compressed latent space is crucial for a variety of applications, including generative modeling, music information retrieval (MIR), and audio compression. Generative models trained on latent representations of audio can be significantly more efficient than models trained directly on the data space, especially considering the high dimensionality of high-sample rate waveform samples. Additionally, a well-designed latent space can facilitate downstream MIR tasks by including musically relevant features in low-dimensional embeddings. However, existing state-of-the-art audio autoencoders often present limitations, such as a multi-stage training process, the use of an unstable adversarial objective that requires multiple discriminators, and slow iterative sampling to reconstruct audio waveforms. 
% The autoencoder used in Musika \cite{pasini_musika_2022} requires a two-stage architecture and does not allow for a fully end-to-end training process, while the diffusion-based autoencoder introduced in Mo\^usai \cite{schneider_mousai_2023} requires multiple sampling steps to reconstruct the original audio, which can be excessively slow and impractical. Both autoencoders also suffer from low reconstruction quality, which is not desirable in professional music production applications. Additionally, autoencoders based on Residual Vector Quantization (RVQ) \cite{vqvae,vqvae2,zeghidour_soundstream_2022,defossez_high_2022,kumar_high-fidelity_2023} compress audio into discrete tokens, which facilitates the training of autoregressive models \cite{copet_simple_2023, jukebox, agostinelli_musiclm_2023}, but it can be challenging to use for training other generative models such as Generative Adversarial Networks (GANs, \cite{gan}) or diffusion models \cite{diffusion_original, song_score-based_2020, ho_denoising_2020}.

In this work, we introduce Music2Latent, a novel consistency autoencoder that encodes audio samples into a continuous latent space with a high compression ratio. Music2Latent is trained fully end-to-end using a single consistency loss function, making it easier to train than many existing audio autoencoders that require a careful balance between multiple competing loss terms \cite{pasini_musika_2022, zeghidour_soundstream_2022, defossez_high_2022, kumar_high-fidelity_2023}. Additionally, considering the underlying consistency model \cite{song_consistency_2023, improvedconsistencysong}, Music2Latent can reconstruct samples with high fidelity in a single step, enabling fast and efficient decoding. We evaluate Music2Latent on audio compression metrics, that measure the discrepancy between input and reconstructed samples, and on audio quality metrics, that establish the general audio quality of the reconstructions. Despite not being the primary focus of our model, we also investigate the downstream performance of encoded representations on standard Music Information Retrieval (MIR) tasks. Our experiments demonstrate that Music2Latent reconstructs samples more accurately and with higher audio quality compared to existing continuous autoencoder baselines while providing comparable or better performance on downstream tasks.
Our contributions are as follows:
\begin{compactitem}
\item We introduce Music2Latent, a consistency autoencoder that encodes waveforms into a continuous latent space with a 4096x time compression ratio.
\item We show how it is possible to achieve high-quality reconstructions with a fully end-to-end training process relying on a single loss function.
\item We introduce a frequency-wise self-attention and a frequency learned scaling mechanism, and demonstrate how they improve audio quality.
\item We demonstrate that Music2Latent surpasses existing continuous autoencoder models in terms of reconstruction accuracy and audio quality while achieving competitive performance on downstream MIR tasks.
\end{compactitem}

To the best of our knowledge, we are the first to successfully use consistency training in the music and audio field, and we are the first across all fields to successfully train an end-to-end consistency autoencoder model.
% \stefan{besides several tricks, is there a more general explanation why a consistency autoencoder leads to higher quality output with highly compressed representations? E.g., is it due to stochasticity? When further reducing the latent dims, would the audio quality be still good, but the reconstruction less faithful? If yes, that could lead to a nice trade-off between encoding size and faithfulness with always guaranteed high-quality audio. If my assumptions are true, it would be worthwhile mentioning that in the introduction.}

\section{Related Work}
% Music2Latent builds upon works on autoencoders for latent generative models and on consistency models. In this section, we provide a brief introduction to these fields.
\subsection{Autoencoders for Latent Generative Modeling}
% Autoencoders have emerged as a powerful tool for learning latent representations of data. 
Several autoencoder approaches have been explored in both the image and audio domains.
% \subsubsection{Image Domain}
\vspace{1mm}
\\
\textbf{Image Domain}: 
Vector Quantized Variational Autoencoders (VQ-VAE) \cite{vqvae} introduced the concept of learning discrete latent representations of images through vector quantization. VQ-VAE-2 \cite{vqvae2} extended this approach to hierarchical codebooks, enabling the generation of realistic images using autoregressive models trained on the learned discrete latent codes.
Vector Quantized Generative Adversarial Networks (VQGAN) \cite{vqgan} combine the VQ-VAE framework with adversarial training, incorporating a discriminator network to improve the perceptual quality of generated images.
Latent Diffusion Models (LDMs) \cite{rombach_high-resolution_2022} leverage diffusion models trained on the latent space of a pre-trained autoencoder. By operating on a compressed representation of the data, LDMs achieve high-quality image synthesis with reduced computational requirements compared to pixel-based diffusion models.
Diffusion autoencoders \cite{diffae} combine a learnable encoder with a diffusion model as the decoder, aiming to learn a meaningful and decodable representation of images in a fully end-to-end manner. However, they still require a slow iterative sampling process to reconstruct samples.
% \subsubsection{Audio Domain}
\vspace{1mm}
\\
\textbf{Audio Domain}: 
The audio autoencoder proposed in the Musika music generation system \cite{pasini_musika_2022} encodes audio into a continuous latent space by reconstructing the magnitude and phase components of a spectrogram. While Musika achieves fast inference, it requires a two-stage training process combined with an unstable adversarial objective.
Mo\^usai introduces a diffusion autoencoder \cite{schneider_mousai_2023} to learn a compressed invertible audio representation. However, it requires multiple sampling steps for reconstruction.
Several audio autoencoders employ Residual Vector Quantization (RVQ) to learn discrete latent representations. Examples include SoundStream \cite{zeghidour_soundstream_2022}, EnCodec \cite{defossez_high_2022}, and Descript Audio Codec (DAC) \cite{kumar_high-fidelity_2023}. These models are well-suited for training autoregressive models on the latent representations but are less suitable for other generative models such as diffusion, consistency, or GAN-based methods. They also generally produce (discrete) representations at a significantly lower time compression ratio than continuous models, and are thus not directly comparable to our work.

\subsection{Consistency Models}
Consistency models \cite{song_consistency_2023, improvedconsistencysong} offer a novel approach for efficient generative modeling by learning a mapping from any point on a diffusion trajectory to the trajectory's starting point. They have been successfully applied to image generation tasks \cite{lcm}, achieving high-quality results with single-step sampling.
The application of consistency models to audio generation is still relatively unexplored. CoMoSpeech \cite{comospeech} explores consistency distillation for speech synthesis, but it requires a pre-trained diffusion model to be trained.

\begin{figure*}[t]
\centering
\includegraphics[width=0.73\textwidth]{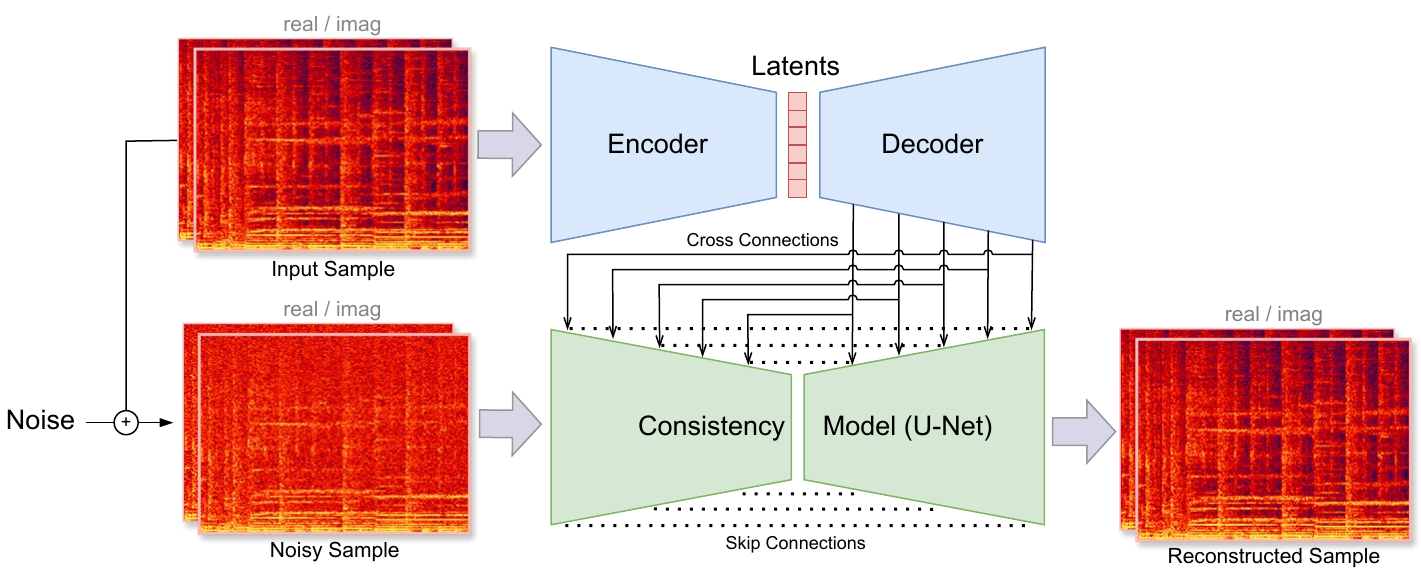}
\caption{Training process of Music2Latent. The input sample is first encoded into a sequence of latent vectors. The latents are then upsampled with a decoder model. The consistency model is trained via consistency training, with an additional information leakage coming from the cross connections.}
\label{fig:training}
\end{figure*}

% \vspace{-1mm}

\vspace{-4mm}
\section{Background}\label{sec:background}
% Music2Latent leverages consistency models \cite{song_consistency_2023, improvedconsistencysong}, which, if trained in isolation and not through distillation, can be considered as a novel class of generative models. 

\subsection{Consistency Models}\label{subsec:consistency_models}
Consistency models represent a novel family of generative models capable of producing high-quality samples in a single step, without the need for adversarial training or iterative sampling. They are grounded in the probability flow ordinary differential equation (ODE) introduced by \cite{song_denoising_2021}:
\begin{equation}
\frac{dx}{d\sigma} = -\sigma\nabla_x \log p_\sigma(x), \quad \sigma \in [\sigma_{\text{min}}, \sigma_{\text{max}}]
\end{equation}
Here, $p_\sigma(x)$ represents the perturbed data distribution obtained by adding Gaussian noise with zero mean and standard deviation $\sigma$ to the original data distribution $p_\text{data}(x)$. The term $\nabla_x \log p_\sigma(x)$ is known as the score function, which plays a crucial role in score-based generative models \cite{song_estimating_gradients, song_score-based_2020, song_improved_score}.
The probability flow ODE establishes a bijective mapping between a noisy data sample $x_\sigma \sim p_\sigma(x)$ and $x_{\sigma_{\text{min}}} \sim p_{\sigma_{\text{min}}}(x) \approx x \sim p_\text{data}(x)$. This mapping, denoted as $f(x_\sigma, \sigma) \mapsto x_{\sigma_{\text{min}}}$, is termed the consistency function, which satisfies the boundary condition $f(x_{\sigma_{\text{min}}}, \sigma_{\text{min}}) = x_{\sigma_{\text{min}}}$.
A consistency model $f_\theta(x_\sigma, \sigma)$ is a neural network trained to approximate the consistency function $f(x_\sigma, \sigma)$. To meet the boundary condition, consistency models are parameterised as:
\begin{equation}\label{eq:consistency_parameterization}
f_\theta(x_\sigma, \sigma) = c_{\text{skip}}(\sigma)x_\sigma + c_{\text{out}}(\sigma)F_\theta(x_\sigma, \sigma)
\end{equation}
where $F_\theta(x_\sigma, \sigma)$ is a free-form neural network, and $c_{\text{skip}}(\sigma)$ and $c_{\text{out}}(\sigma)$ are differentiable functions such that $c_{\text{skip}}(\sigma_{\text{min}}) = 1$ and $c_{\text{out}}(\sigma_{\text{min}}) = 0$.

Consistency models can be trained using either consistency distillation (CD) or consistency training (CT). CD requires pre-training a diffusion model to estimate the score function $\nabla_x \log p_\sigma(x)$ via score matching \cite{song_sliced_2019}. CT, on the other hand, allows training consistency models in isolation and is the method that is considered in this work.

\subsection{Consistency Training}\label{subsec:consistency_training}
In consistency training, the probability flow ODE is discretised using a sequence of noise levels $\sigma_{\text{min}} = \sigma_1 < \sigma_2 < \cdots < \sigma_N = \sigma_{\text{max}}$. The consistency model $f_\theta(x_\sigma, \sigma)$ is then trained by minimising the following consistency training loss over $\theta$:
\begin{equation}\label{eq:consistency_training_loss}
\mathcal{L}_\text{CT} = \mathbb{E} \left[ \lambda(\sigma_i, \sigma_{i+1}) d\left(f_\theta(x_{\sigma_{i+1}}, \sigma_{i+1}), f_{\theta^-}(x_{\sigma_i}, \sigma_i)\right) \right]
\end{equation}
where $d(x, y)$ is a metric function such as mean squared error and $\lambda(\sigma_i, \sigma_{i+1})$ is a noise level-dependent loss scaling.
In the above equations, $f_\theta$ and $f_{\theta^-}$ are referred to as the student network and the teacher network respectively. The teacher's parameters $\theta^-$ are obtained by applying a stop-gradient operation to the student's parameters $\theta$ during training:
\begin{equation}
\theta^- \gets \text{stopgrad}(\theta)
\end{equation}
After training, the consistency model $f_\theta(x, \sigma)$ can directly generate a sample $x$ by starting with $z \sim \mathcal{N}(0, I)$ and computing $x = f_\theta(\sigma_\text{max} z, \sigma_\text{max})$. This enables efficient one-step sampling, a key advantage of consistency models over diffusion models.
% Consistency models also support multi-step sampling, enabling a trade-off between sample quality and computational efficiency. 
% For a sequence of indices $1 = i_1 < i_2 < \cdots < i_K = N$, one can start by sampling $x_K \sim \mathcal{N}(0, \sigma_\text{max}^2)$ and then iteratively compute:
% \begin{equation}
% x_k \gets f_\theta(x_{k+1}, \sigma_{i_{k+1}}) + \sqrt{\sigma_{i_k}^2 - \sigma_\text{min}^2} z_k
% \end{equation}
% where $z_k \sim \mathcal{N}(0, I)$. The resulting sample $x_1$ approximates the data distribution $p_\text{data}(x)$.

\vspace{-3mm}
\section{Music2Latent}

In the following sections, we provide a detailed explanation of the audio representation, architecture, and training framework underlying Music2Latent.

\subsection{Audio Representation}\label{subsec:audio_representation}
Music2Latent utilises complex-valued STFT spectrograms as the representation of waveform audio. This choice is motivated by several factors. First, previous works \cite{drumgan, comparing} have demonstrated the effectiveness of complex spectrograms in capturing the intricate structure of audio signals and enabling the generation of high-fidelity audio. Second, 2-dimensional spectrograms allow for the direct application of UNet architectures \cite{unet} that have been successfully used in the image domain with diffusion and consistency models.
However, the distribution of values across different frequencies in a STFT spectrogram can vary significantly, with substantially higher magnitudes in low frequencies compared to high frequencies. This can hinder the ability of the model to accurately reconstruct all frequency components, as the learning signal for high frequencies may be overshadowed by the stronger signal from lower frequencies. To address this issue, we apply the amplitude transformation proposed in \cite{speech_enhancement} and later used in \cite{zhu2023edmsound} which scales up lower energy components in the spectrogram:
\begin{equation}
\tilde{c} = \beta |c|^\alpha e^{i \angle(c)}
\end{equation}
where $c$ is the original complex STFT coefficient, $\tilde{c}$ is the transformed coefficient, $\alpha \in (0, 1]$ is a compression exponent that emphasizes lower-energy frequency components, $\angle(c)$ represents the phase angle of $c$, and $\beta \in \mathbb{R}^{+}$ is a scaling factor to normalize amplitudes within a desired range (e.g., $[0, 1]$). This transformation ensures that the model receives a more balanced representation of the audio signal, facilitating accurate reconstruction across all frequencies. We consider the complex STFT spectrogram as a 2-channel representation, with each channel representing real and imaginary components respectively.

\subsection{Architecture}
The architecture of Music2Latent consists of an encoder, a decoder, and a consistency model. 
% We provide a detailed explanation of these architectural components in the following sections.
% \subsubsection{Encoder}
\vspace{1mm}
\\
\textbf{Encoder}: 
The encoder receives as input the audio sample in the form of an STFT spectrogram with real and imaginary components in each channel. It then gradually downsamples the feature maps along the time axis and outputs a sequence of latent vectors with dimensionality $d_{lat}$. Instead of being trained with a VAE objective \cite{vae, rombach_high-resolution_2022} to keep the distribution of latent values under control, the latent encodings of the model are kept in the $(-1,1)$ range using a $tanh$ activation function, which was proven to be a successful approach in previous works for downstream latent generative modeling tasks \cite{pasini_musika_2022, schneider_mousai_2023}.
% \subsubsection{Decoder}
\vspace{1mm}
\\
\textbf{Decoder}: 
The decoder mirrors the encoder architecture but performs upsampling instead of downsampling. The decoder takes as input a sequence of latent vectors from the encoder and progressively upsamples them to match the dimensionality of the feature maps of the consistency model. The only purpose of the decoder is to ensure that the conditioning information from the latent encodings is available to the consistency model at all levels of its architecture (the reason for this architectural choice is provided in the next section).
% \subsubsection{Consistency Model}
\vspace{1mm}
\\
\textbf{Consistency Model}: 
The consistency model uses a UNet architecture with a downsampling branch and an upsampling branch connected via additive skip connections. The output of the decoder at each upsampling layer is also added to the corresponding layer of the consistency model. This provides cross connections that allow the consistency model to directly access the conditioning information about the sample it is attempting to reconstruct at all levels of its architecture. This design choice is crucial for single-step reconstruction, as it ensures that the model has access to the necessary information to accurately reconstruct the target sample from the very beginning of the UNet architecture.
% \subsubsection{Adaptive Frequency Scaling}
\vspace{1mm}
\\
\textbf{Adaptive Frequency Scaling}: 
The distribution of values along the frequency axis in the input spectrograms changes significantly with respect to the noise level $\sigma$. Specifically, when $\sigma$ is close to $\sigma_{min}$, the magnitudes at low frequencies are on average much higher than the ones at high frequencies, while with $\sigma$ approaching $\sigma_{max}$, there is an equal distribution of values across all frequencies since the sample is pure noise. To address this, we introduce a frequency-wise scaling mechanism that adaptively scales the input and output of the consistency model based on the current noise level.
Specifically, we employ a Multi-Layer Perceptron (MLP) that takes as input the noise level $\sigma$ in the form of a sinusoidal embedding \cite{transformer} and outputs a scaling factor for each frequency bin:
\begin{equation}
s_f(\sigma) = \text{MLP}(\sigma),
\end{equation}
where $s_f(\sigma) \in \mathbb{R}^F$ is a vector of scaling factors, one for each of the $F$ frequency bins of the noisy spectrogram. We calculate different scaling factors to scale both the input $x_{\sigma}$ and the output of the consistency model $F_\theta(x_\sigma)$ as follows:
\begin{equation}
\tilde{x}_{\sigma} = x_{\sigma} \odot s_{f, \text{in}}(\sigma) \quad \tilde{F}_\theta(x_\sigma) = F_\theta(x_\sigma) \odot s_{f, \text{out}}(\sigma)
\end{equation}
where $\odot$ denotes element-wise multiplication.
% \subsubsection{Frequency-wise self-attention}
\vspace{1mm}
\\
\textbf{Frequency-wise self-attention}: 
To capture long-range dependencies within the frequency domain while keeping a memory footprint that scales linearly with the time axis, Music2Latent employs frequency-wise self-attention. This mechanism allows the model to attend to information from all frequency bins at a given time step, enabling it to learn complex relationships between different frequency components. Considering that only the time dimension of the input can vary at inference time, using frequency-wise attention compared to full self-attention does not incur in a memory requirement that scales quadratically with time.
After computing the query $Q$, key $K$, and value $V$ via linear projections of the input features, we calculate the attention matrix $A$ by performing an outer product on individual timesteps $t$:
\begin{equation}
A_t = \text{softmax}\left(\frac{Q_t K_t^T}{\sqrt{d}}\right)
\end{equation}
where $d$ is the channel dimension, and after concatenating the attention weights from all timesteps together we have $A \in \mathbb{R}^{T \times F \times F}$. The softmax operation is then applied across the frequency dimension, ensuring that the attention weights for each frequency bin sum to one.

\begin{table*}[t]
\begin{footnotesize}
\centering
\begin{tabular}{lcccccccc}
\toprule
& \multicolumn{2}{c}{MagnaTagATune} & \multicolumn{2}{c}{Beatport} & \multicolumn{2}{c}{TinySOL-pitchclass} & \multicolumn{2}{c}{TinySOL-instrument} \\
& AUC-ROC & AUC-PR & Micro Acc. & Macro Acc. & Micro F1 & Macro F1 & Micro F1 & Macro F1 \\
\midrule
Musika & 84.8 & 32.9 & 45.2 & 41.0 & 93.5 & 93.4 & \underline{93.3} & \underline{84.5} \\
LatMusic & 85.9 & 34.9 & 37.4 & 30.2 & 88.9 & 88.8 & 92.6 & 80.7 \\
Mo\^usai\_v2 & 86.2 & 35.4 & 48.2 & 42.0 & 95.1 & 95.1 & 82.8 & 68.6 \\
Mo\^usai\_v3 & 85.8 & 34.5 & 39.8 & 31.9 & 95.5 & 95.6 & 93.1 & 82.3 \\
\textbf{Music2Latent} & \underline{88.6} & \underline{40.0} & \underline{\textbf{65.5}} & \underline{\textbf{60.1}} & \underline{\textbf{99.8}} & \underline{\textbf{99.8}} & 92.6 & 81.0 \\
\midrule
MusiCNN-MSD & 87.6 & 37.5 & 13.5 & 7.3 & 17.2 & 15.7 & 68.2 & 60.8 \\
CLMR & 89.9 & 42.6 & 13.9 & 7.8 & 16.8 & 16.2 & 93.5 & 89.7 \\
MERT-v1-95M & \textbf{90.8} & \textbf{44.9} & 50.7 & 44.3 & 98.3 & 98.3 & \textbf{97.1} & \textbf{95.8} \\
\bottomrule
\end{tabular}
\caption{Downstream task performance on MagnaTagATune (autotagging), Beatport (key estimation), TinySOL (pitch and instrument classification). Best results among autoencoder baselines are underlined.}
\label{tab:downstream}
\end{footnotesize}
\vspace{-0.3cm}
\end{table*}

\subsection{Training Process}
Music2Latent is trained using the consistency training (CT) objective \cite{song_consistency_2023, improvedconsistencysong}. As described in Sec.~\ref{subsec:consistency_training}, the objective minimizes the discrepancy between the outputs of the consistency model at adjacent noise levels $\sigma_i$ and $\sigma_{i+1}$.
As for the distance metric in the consistency training loss function (Eq. \ref{eq:consistency_training_loss}), we use the Pseudo-Huber loss function \cite{huber_loss} which smoothly transitions from the $\ell_1$ to the squared $\ell_2$ metrics:
\begin{equation}
d(x, y) = \sqrt{|x - y|^2 + c^2} - c,
\end{equation}
where $c$ is a hyperparameter that controls the transition. In \cite{improvedconsistencysong}, it was shown that for image generation with consistency models, this loss provides smoother gradients during training and performs substantially better compared to the more common squared $\ell_2$ loss.
The consistency model is parameterised as described in Eq. \ref{eq:consistency_parameterization}, with the exception that in addition to providing as input the noisy sample $x_{\sigma}$, we allow for information leakage of the clean sample $x$ through the features $\bm{y}_x$ provided by the decoder via cross connections:
% \begin{equation}
% \begin{aligned}\label{eq:training}
% \text{lat}_x &= \text{Enc}_\theta(x) \tab \tab \\
% \bm{y}_x &= \text{Dec}_\theta(\text{lat}_x) \\
% f_\theta(x_\sigma, \sigma, \bm{y}_x) &= c_{\text{skip}}(\sigma)x_\sigma + c_{\text{out}}(\sigma)F_\theta(x_\sigma, \sigma, \bm{y}_x)
% \end{aligned}
% \end{equation}
\begin{equation}
\begin{alignedat}{2}\label{eq:training}
\text{lat}_x = \text{Enc}_\theta(x) &\quad\bm{y}_x = \text{Dec}_\theta(\text{lat}_x) \\
f_\theta(x_\sigma, \sigma, \bm{y}_x) = c_{\text{skip}}(\sigma)x_\sigma &+ c_{\text{out}}(\sigma)F_\theta(x_\sigma, \sigma, \bm{y}_x)
\end{alignedat}
\end{equation}

which results in the following consistency loss that is used to train the system fully end-to-end:
\begin{equation}\label{eq:consistency_training_loss_new}
\mathcal{L} = \mathbb{E} \left[ \lambda(\sigma_i, \sigma_{i+1}) d\left(f_\theta(x_{\sigma_{i+1}}, \sigma_{i+1}, \bm{y}_x), f_{\theta^-}(x_{\sigma_i}, \sigma_i, \bm{y}_x)\right) \right]
\end{equation}
With respect to the noise level-dependent loss scaling $\lambda(\sigma_i, \sigma_{i+1})$, we follow \cite{improvedconsistencysong} and use:
\begin{equation}
   \lambda(\sigma_i, \sigma_{i+1}) = \frac{1}{\sigma_{i+1}-\sigma_i}
\end{equation}
which assigns a higher weight to the loss when there is a small gap between consecutive noise levels.
We also adopt the lognormal sampling of $\sigma$ introduced by \cite{karras_elucidating_2022} and adopted for consistency training by \cite{improvedconsistencysong} to focus training on a more relevant range of noise levels.
% \subsubsection{Continuous Noise Levels}
\vspace{1mm}
\\
\textbf{Continuous Noise Levels}: 
Unlike the formulation presented in previous consistency model literature \cite{song_consistency_2023, improvedconsistencysong}, which use a discrete set of noise levels for training, Music2Latent employs a continuous noise schedule. This change is inspired by recent state-of-the-art diffusion models which notably sample noise levels from a continuous distribution \cite{karras_elucidating_2022}. Parallel work on improving the performance of consistency models also demonstrates how employing a continuous noise schedule improves results compared to the original discrete schedule \cite{easy_consistency_models}.
Specifically, we use an exponential schedule during training to determine the step size between consecutive noise levels used for the consistency loss:
\begin{equation}
\Delta t_k = \Delta t_0^{\frac{k}{K}\left(e_K-1\right)+1}
\end{equation}
where $\Delta t_k$ is the step size at training iteration $k$, $\Delta t_0$ is the initial step size at iteration 0, and $e_K$ is the exponent at final iteration $K$. This schedule ensures that the step size decreases exponentially as training progresses, allowing the model to gradually learn finer details of the data distribution.
In order to calculate $\sigma_i$ and $\sigma_{i+1}$, we first sample a timestep $t_{i+1} \in \left[0,1\right]$ with the sampling weights given by the lognormal distribution, and calculate the adjacent timestep $t_i=\max(t_{i+1}-\Delta t_k, 0)$. Finally we calculate $\sigma_i$ using the time step-to-noise level mapping from \cite{karras_elucidating_2022}:
\begin{equation}
    \sigma_i = \left( \sigma_{\text{min}}^{\frac{1}{\rho}} + t_i \left( \sigma_{\text{max}}^{\frac{1}{\rho}} - \sigma_{\text{min}}^{\frac{1}{\rho}} \right) \right)^\rho
\end{equation}
where $\rho=7$. We use the same mapping to calculate $\sigma_{i+1}$.

% \begin{table*}[t]
% \begin{footnotesize}
% \centering
% \begin{tabular}{lcccccccc}
% \toprule
% & \multicolumn{2}{c}{MagnaTagATune} & \multicolumn{2}{c}{Beatport} & \multicolumn{2}{c}{TinySOL-pitchclass} & \multicolumn{2}{c}{TinySOL-instrument} \\
% & AUC-ROC & AUC-PR & Micro Acc. & Macro Acc. & Micro F1 & Macro F1 & Micro F1 & Macro F1 \\
% \midrule
% Musika & 84.8 & 32.9 & 45.2 & 41.0 & 93.5 & 93.4 & \textbf{93.3} & \textbf{84.5} \\
% LatMusic & 85.9 & 34.9 & 37.4 & 30.2 & 88.9 & 88.8 & 92.6 & 80.7 \\
% Mo\^usai\_v2 & 86.2 & 35.4 & 48.2 & 42.0 & 95.1 & 95.1 & 82.8 & 68.6 \\
% Mo\^usai\_v3 & 85.8 & 34.5 & 39.8 & 31.9 & 95.5 & 95.6 & 93.1 & 82.3 \\
% \textbf{Music2Latent} & \textbf{88.6} & \textbf{40.0} & \textbf{65.5} & \textbf{60.1} & \textbf{99.8} & \textbf{99.8} & 92.6 & 81.0 \\
% \midrule
% MusiCNN-MSD & 87.6 & 37.5 & 13.5 & 7.3 & 17.2 & 15.7 & 68.2 & 60.8 \\
% CLMR & 89.9 & 42.6 & 13.9 & 7.8 & 16.8 & 16.2 & 93.5 & 89.7 \\
% MERT-v1-95M & 90.8 & 44.9 & 50.7 & 44.3 & 98.3 & 98.3 & 97.1 & 95.8 \\
% \bottomrule
% \end{tabular}
% \caption{\footnotesize{Downstream task performance on MagnaTagATune, Beatport, TinySOL-pitchclass, and TinySOL-instrument.}}
% \label{tab:downstream}
% \end{footnotesize}
% \end{table*}

\subsection{Implementation Details}\label{subsec:implementation}
With respect to the UNet architecture of the consistency model, we use the NCSN++ architecture introduced in \cite{song_score-based_2020}, which consists of convolutional residual blocks with 3x3 kernels, Swish activation function \cite{swish} and Group Normalisation layers. The same residual blocks are used in both the encoder and decoder. We use sinusoidal embeddings to encode the noise level, using $\frac{\log(\sigma)}{4}$ as the input. The skip connections between the downsampling and the upsampling branches of the UNet are added instead of being concatenated, as recent works on diffusion models \cite{sd3} show that addition provides better performance. Consequently, the cross connections from the decoder are also added to the corresponding UNet features, following a linear projection layer. In the encoder, before the final bottleneck layer with a $tanh$ activation function, used to constrain the latent encodings to the (-1,1) range, the 2D features are reshaped into 1D features by flattening the frequency dimension into the channel dimension, and a series of 4 residual blocks with 1D convolutions with kernel size of 3 are used. We choose $d_{lat}=64$, which results in a $4096x$ time compression ratio and a $64x$ total compression ratio. The decoder perfectly mirrors the architecture of the encoder, while not receiving any incoming skip connections, since all the information necessary to reconstruct the clean input sample must be contained in the latent encodings. For the consistency model and encoder/decoder models we use 5 levels corresponding to 4 upsampling/downsampling operations, and in each level we use 2 residual blocks for the consistency model, and 1 residual block for the encoder and decoder. The base channels for all models are set to 64 and the channel multiplier for each of the 5 levels is set to $[1,2,4,4,4]$ for all models. We use $512$ channels for the 1D convolutional blocks in the encoder and decoder. We use frequency-wise self-attention layers with $4$ heads in the 3 last levels for all models, in order not to use it with higher frequency dimensions. The channels used for sinusoidal embeddings and the MLPs used for both noise level embeddings and frequency scalings are set to $256$. The model has $\sim 58$ million parameters.
The consistency training framework follows the same implementation of \cite{improvedconsistencysong} with respect to the scaling factors $c_{\text{in}}, c_{\text{skip}}, c_{\text{out}}$, the parameter $c$ for the pseudo-Huber loss function, the minimum and maximum noise parameters $\sigma_{\text{min}}, \sigma_{\text{max}}$, the standard deviation of the data samples $\sigma_{\text{data}}$, and the lognormal distribution values of $P_{\text{mean}}, P_{\text{std}}$. Regarding the input STFT spectrograms, we extract them using $\text{hop}=512, \text{window}=4 \cdot \text{hop}$ and we transform them using the formula presented in Sec.~\ref{subsec:audio_representation}, with $\alpha=0.65, \beta=0.35$. Regarding the step size schedule for the continuous noise levels, we choose $\Delta t_0=0.1$ and $e_K=3$.
We train the model on waveforms of $34,304$ samples, which correspond to $\sim0.78\, \text{s}$ of $44.1\, \text{kHz}$ audio. The model thus produces latent representations of $44.1\, \text{kHz}$ audio at a sampling rate of $\sim 11\, \text{Hz}$. We use a batch size of $16$ and train for $K=800k$ iterations using the RAdam optimizer \cite{radam} with $\text{lr}_0=1e^{-4}, \beta_1=0.9, \beta_2=0.999$. We use a cosine learning rate decay with $\text{lr}_K=1e^{-6}$ and we keep an Exponential Moving Average (EMA) of the parameters of all models with a momentum of $0.9999$. Training takes $\sim 5$ days on a single RTX 3090 GPU.

\begin{figure}[h]
\centering
\includegraphics[width=0.45\textwidth]{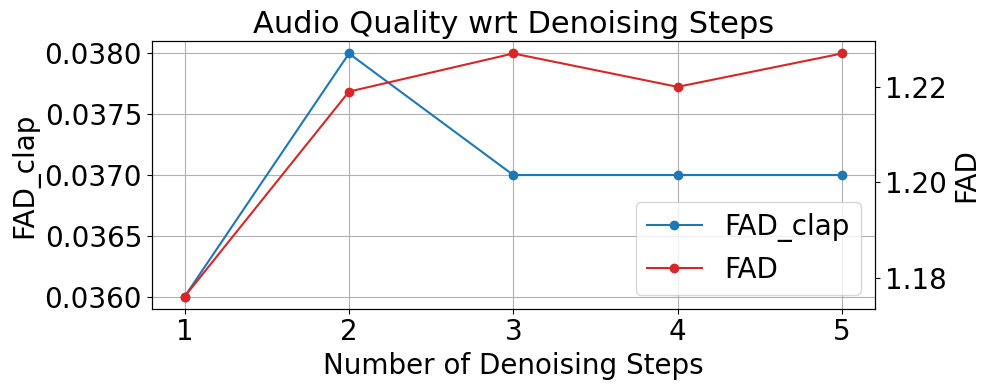}
\caption{Audio quality of reconstructed samples with respect to the number of denoising steps of the consistency model.}
\label{fig:qualitytradeoff}
\end{figure}

\vspace{-3mm}
\section{Experiments and Results}

\subsection{Datasets}
We train the model on MTG Jamendo \cite{mtg_jamendo} and on the clean speech segments from DNS Challenge 4 \cite{dns}, sampling from each dataset with equal probability. We keep the original sample rates of $44.1\, \text{kHz}$ and $48\, \text{kHz}$. We include speech in the training data to both improve the reconstruction of vocal content in music samples, and to make Music2Latent useful also for speech-related tasks. We use MusicCaps \cite{agostinelli_musiclm_2023} as our evaluation dataset.
\subsection{Baselines}
We compare Music2Latent to different audio autoencoders that encode audio samples into a continuous latent space to enable downstream latent generative modeling.  We include the autoencoder introduced in Musika \cite{pasini_musika_2022} and the autoencoder introduced by \cite{bassnet} to train a latent diffusion model for music accompaniment generation (we name this model LatMusic in our comparison). Both models encode audio samples with the same compression ratio of $64x$ as Music2Latent. We also include the diffusion autoencoder introduced in Mo\^usai \cite{schneider_mousai_2023}, which has a compression ratio of $32x$ (Mo\^usai\_v3), and a different autoencoder model that is made available by the authors of Mo\^usai\footnote{\href{https://github.com/archinetai/archisound}{https://github.com/archinetai/archisound}} with a comparable compression ratio of $64x$ (Mo\^usai\_v2).

\subsection{Audio Compression and Quality}

\begin{table}[h]
\begin{footnotesize}
\centering
\begin{tabular}{lcccc}
\toprule
& SI-SDR $\uparrow$ & ViSQOL $\uparrow$ & $\text{FAD}_\text{clap}\downarrow$ & $\text{FAD}\downarrow$ \\
\midrule
Musika & -25.81 & 3.80 & 0.103 & 2.308 \\
LatMusic & -27.32 & \textbf{3.95} & 0.050 & 1.630 \\
Mo\^usai\_v2 & -21.44 & 2.36 & 0.731 & 4.687 \\
Mo\^usai\_v3 & -17.47 & 2.28 & 0.647 & 4.473 \\
\textbf{Music2Latent} & \textbf{-3.85} & 3.84 & \textbf{0.036} & \textbf{1.176} \\
\midrule
\textit{DAC} & \textit{9.48} & \textit{4.21} & \textit{0.041} & \textit{0.966} \\
\bottomrule
\end{tabular}
\caption{Audio compression and quality results.}
\label{tab:audio_quality}
\end{footnotesize}
\end{table}

% \begin{figure}[h]
% \centering
% \includegraphics[width=0.4\textwidth]{figs/qualityvssteps.png}
% \caption{\footnotesize{Audio quality of reconstructed samples with respect to the number of denoising steps of the consistency model.}}
% \label{fig:qualitytradeoff}
% \end{figure}

% \begin{table}[h]
% \begin{footnotesize}
% \centering
% \begin{tabular}{lcccc}
% \toprule
% & SI-SDR $\uparrow$ & ViSQOL $\uparrow$ & $\text{FAD}_\text{clap}\downarrow$ & $\text{FAD}\downarrow$ \\
% \midrule
% Musika & -25.81 & 3.80 & 0.103 & 2.308 \\
% LatMusic & -27.32 & \textbf{3.95} & 0.050 & 1.630 \\
% Mo\^usai\_v2 & -21.44 & 2.36 & 0.731 & 4.687 \\
% Mo\^usai\_v3 & -17.47 & 2.28 & 0.647 & 4.473 \\
% \textbf{Music2Latent} & \textbf{-3.85} & 3.84 & \textbf{0.036} & \textbf{1.176} \\
% \midrule
% DAC & 9.48 & 4.21 & 0.041 & 0.966 \\
% \bottomrule
% \end{tabular}
% \caption{\footnotesize{Audio compression and quality results.}}
% \label{tab:audio_quality}
% \end{footnotesize}
% \end{table}

We adopt the same objective evaluation metrics as in \cite{defossez_high_2022} and use  Scale-Invariant Signal-to-Noise Ratio (SI-SDR) \cite{sisdr} and ViSQOL \cite{visqol, visqolaudio, visqolv3}. SI-SDR is a distance calculated between input and reconstructed waveforms, while ViSQOL estimates a MOS-like score on perceptual quality from the difference between the two signals. Considering that Music2Latent is trained as a generative model, we also use Frechét Audio Distance (FAD \cite{fad}) to evaluate the general audio quality of reconstructed samples without relying on paired samples. In addition to the original FAD implementation, we also evaluate on $\text{FAD}_\text{clap}$ using CLAP \cite{clap} features, which was shown to correlate significantly better with perceived audio quality \cite{fad_correlation}. In Tab. \ref{tab:audio_quality} we show that Music2Latent is competitive with respect to ViSQOL to Musika and LatMusic, while vastly outperforming all baselines on the remaining metrics. Note that all four baselines discard phase information from the input of the autoencoder, which may explain the poor SI-SDR performance. DAC, while not being directly comparable, scores favourably in reconstruction metrics, while matches Music2Latent in terms of audio quality. In Fig. \ref{fig:qualitytradeoff} we also show that the audio quality of reconstructions remains almost constant when using more than a single denoising step. We provide audio samples and additional supplementary material on the accompanying website\footnote{\href{https://sonycslparis.github.io/music2latent-companion/}{https://sonycslparis.github.io/music2latent-companion/}}.

\subsection{Ablation Study}
\begin{table}[h]
\begin{footnotesize}
\centering
\begin{tabular}{lcc}
\toprule
& $\text{FAD}_\text{clap}\downarrow$ & $\text{FAD}\downarrow$ \\
\midrule
Base Model & 0.0563 & 1.808 \\
+ Freq-wise Attention & 0.0547 & 1.710 \\
+ Adaptive Freq Scaling & \textbf{0.0537} & \textbf{1.665} \\
\bottomrule
\end{tabular}
\caption{Ablation study. \textit{Base Model} is trained without frequency-wise attention and adaptive frequency scaling.}
\label{tab:ablation}
\end{footnotesize}
\end{table}

To demonstrate the effectiveness of both frequency-wise attention and learned frequency scaling, we perform an ablation study and report the FAD and $\text{FAD}_\text{clap}$ results in Table \ref{tab:ablation}. With respect to the model with no attention and no scaling, we use channel multipliers $[1,2,4,4,5]$ to roughly match the number of parameters that are lost. All ablated models are trained for $200k$ iterations. The remaining training details are the ones described in Sec.~\ref{subsec:implementation}.

\subsection{Downstream Performance}
Since training representation learning models on compressed audio representations instead of raw data was shown to be a promising approach \cite{repr_calm, repr_encodecmae, repr_pecmae, repr_mert}, our goal is to investigate whether there are well disentangled audio features in the feature space of audio autoencoders. We evaluate downstream performance on MagnaTagATune \cite{magnatagatune} for auto-tagging, Beatport \cite{beatport} for key estimation, and TinySOL \cite{tinysol} for instrument and pitch class classification. For each dataset, we extract the encoder features from the layer with the highest number of output channels from each of the models (after flattening the 2D features for Music2Latent and before the last linear layer for the remaining models), average them along the time axis, and train a 2-layer MLP with $[256,128]$ units. We also show the results obtained by performing the same evaluation on features from the classification model MusiCNN-MSD \cite{musicnn} and well-established representation learning models CLMR \cite{clmr} and MERT-v1-95M \cite{repr_mert} (with averaged features from layers 9 to 12). We extract features from these models following \cite{christos_thesis} and perform all evaluations using the \texttt{mir\_ref} library\footnote{\href{https://github.com/chrispla/mir_ref}{https://github.com/chrispla/mir\_ref}} \cite{mir_ref}. In Tab. \ref{tab:downstream} we show how Music2Latent outperforms autoencoder baselines in almost all tasks, and in the case of key and pitch classification it even outperforms state-of-the-art representation learning models. We hypothesize that the loss is more sensitive to pitch information than timbre content (explaining the weak comparison on TinySOL-instrument to the representation learning models). 
% We leave further analysis of downstream performance as future work.   

\vspace{-3mm}
\section{Conclusion}
In this work we introduced Music2Latent, a consistency autoencoder that efficiently compresses high-dimensional audio waveforms into a continuous latent space. By leveraging consistency training, Music2Latent achieves high-fidelity single-step reconstruction, and enables efficient downstream latent generative modeling.
We propose a learned frequency scaling mechanism to handle varying frequency distributions across diffusion noise levels. Experiments show Music2Latent matches or outperforms baselines in reconstruction accuracy and audio quality, while having comparable or better performance on downstream tasks.
To our knowledge, Music2Latent represents the first successful end-to-end consistency autoencoder. Future work could explore extensions to other modalities and higher compression ratios.
Overall, we believe Music2Latent is a significant contribution to audio generative modeling and representation learning.

\section{Acknowledgements}
This work is supported by the EPSRC UKRI Centre for
Doctoral Training in Artificial Intelligence and Music (EP/S022694/1) and Sony Computer Science Laboratories Paris.
\bibliography{ISMIRtemplate}

% Generated by IEEEtran.bst, version: 1.14 (2015/08/26)
\begin{thebibliography}{10}
\providecommand{\url}[1]{#1}
\csname url@samestyle\endcsname
\providecommand{\newblock}{\relax}
\providecommand{\bibinfo}[2]{#2}
\providecommand{\BIBentrySTDinterwordspacing}{\spaceskip=0pt\relax}
\providecommand{\BIBentryALTinterwordstretchfactor}{4}
\providecommand{\BIBentryALTinterwordspacing}{\spaceskip=\fontdimen2\font plus
\BIBentryALTinterwordstretchfactor\fontdimen3\font minus \fontdimen4\font\relax}
\providecommand{\BIBforeignlanguage}[2]{{%
\expandafter\ifx\csname l@#1\endcsname\relax
\typeout{** WARNING: IEEEtran.bst: No hyphenation pattern has been}%
\typeout{** loaded for the language `#1'. Using the pattern for}%
\typeout{** the default language instead.}%
\else
\language=\csname l@#1\endcsname
\fi
#2}}
\providecommand{\BIBdecl}{\relax}
\BIBdecl

\bibitem{pasini_musika_2022}
M.~Pasini and J.~Schlüter, ``Musika! {Fast} {Infinite} {Waveform} {Music} {Generation},'' in \emph{Proceedings of the 23rd {International} {Society} for {Music} {Information} {Retrieval} {Conference}, {ISMIR} 2022, {Bengaluru}, {India}, {December} 4-8, 2022}, 2022, pp. 543--550.

\bibitem{zeghidour_soundstream_2022}
N.~Zeghidour, A.~Luebs \emph{et~al.}, ``{SoundStream}: {An} {End}-to-{End} {Neural} {Audio} {Codec},'' \emph{IEEE ACM Trans. Audio Speech Lang. Process.}, vol.~30, pp. 495--507, 2022.

\bibitem{defossez_high_2022}
A.~Défossez, J.~Copet \emph{et~al.}, ``High {Fidelity} {Neural} {Audio} {Compression},'' Oct. 2022, arXiv:2210.13438 [cs, eess, stat].

\bibitem{kumar_high-fidelity_2023}
R.~Kumar, P.~Seetharaman \emph{et~al.}, ``High-{Fidelity} {Audio} {Compression} with {Improved} {RVQGAN},'' Jun. 2023, arXiv:2306.06546 [cs, eess].

\bibitem{song_consistency_2023}
Y.~Song, P.~Dhariwal \emph{et~al.}, ``Consistency {Models},'' May 2023, arXiv:2303.01469 [cs, stat].

\bibitem{improvedconsistencysong}
Y.~Song and P.~Dhariwal, ``Improved techniques for training consistency models,'' \emph{arXiv preprint arXiv:2310.14189}, 2023.

\bibitem{vqvae}
A.~van~den Oord, O.~Vinyals \emph{et~al.}, ``Neural discrete representation learning,'' in \emph{Advances in Neural Information Processing Systems 30}, Dec. 2017, pp. 6306--6315.

\bibitem{vqvae2}
A.~Razavi, A.~van~den Oord \emph{et~al.}, ``Generating diverse high-fidelity images with {VQ-VAE-2},'' in \emph{Advances in Neural Information Processing Systems 32}, Dec. 2019, pp. 14\,837--14\,847.

\bibitem{vqgan}
P.~Esser, R.~Rombach \emph{et~al.}, ``Taming transformers for high-resolution image synthesis,'' in \emph{{IEEE} Conference on Computer Vision and Pattern Recognition (CVPR)}.\hskip 1em plus 0.5em minus 0.4em\relax Computer Vision Foundation / {IEEE}, Jun. 2021, pp. 12\,873--12\,883.

\bibitem{rombach_high-resolution_2022}
R.~Rombach, A.~Blattmann \emph{et~al.}, ``High-resolution image synthesis with latent diffusion models,'' in \emph{Proceedings of the {IEEE}/{CVF} {Conference} on {Computer} {Vision} and {Pattern} {Recognition}}, 2022, pp. 10\,684--10\,695.

\bibitem{diffae}
K.~Preechakul, N.~Chatthee \emph{et~al.}, ``Diffusion autoencoders: Toward a meaningful and decodable representation,'' in \emph{{IEEE/CVF} Conference on Computer Vision and Pattern Recognition, {CVPR} 2022, New Orleans, LA, USA, June 18-24, 2022}.\hskip 1em plus 0.5em minus 0.4em\relax {IEEE}, 2022, pp. 10\,609--10\,619.

\bibitem{schneider_mousai_2023}
F.~Schneider, Z.~Jin \emph{et~al.}, ``Mo{\textbackslash}{\textasciicircum}usai: {Text}-to-{Music} {Generation} with {Long}-{Context} {Latent} {Diffusion},'' Jan. 2023, arXiv:2301.11757 [cs, eess].

\bibitem{lcm}
S.~Luo, Y.~Tan, L.~Huang, J.~Li, and H.~Zhao, ``Latent consistency models: Synthesizing high-resolution images with few-step inference,'' \emph{arXiv preprint arXiv:2310.04378}, 2023.

\bibitem{comospeech}
Z.~Ye, W.~Xue \emph{et~al.}, ``Comospeech: One-step speech and singing voice synthesis via consistency model,'' in \emph{Proceedings of the 31st {ACM} International Conference on Multimedia, {MM} 2023, Ottawa, ON, Canada, 29 October 2023- 3 November 2023}.\hskip 1em plus 0.5em minus 0.4em\relax {ACM}, 2023, pp. 1831--1839.

\bibitem{song_denoising_2021}
J.~Song, C.~Meng \emph{et~al.}, ``Denoising {Diffusion} {Implicit} {Models},'' in \emph{9th {International} {Conference} on {Learning} {Representations}, {ICLR} 2021, {Virtual} {Event}, {Austria}, {May} 3-7, 2021}.\hskip 1em plus 0.5em minus 0.4em\relax OpenReview.net, 2021.

\bibitem{song_estimating_gradients}
Y.~Song and S.~Ermon, ``Generative modeling by estimating gradients of the data distribution,'' in \emph{Advances in Neural Information Processing Systems 32: Annual Conference on Neural Information Processing Systems 2019, NeurIPS 2019, December 8-14, 2019, Vancouver, BC, Canada}, 2019, pp. 11\,895--11\,907.

\bibitem{song_score-based_2020}
Y.~Song, J.~Sohl-Dickstein \emph{et~al.}, ``Score-based generative modeling through stochastic differential equations,'' \emph{arXiv preprint arXiv:2011.13456}, 2020.

\bibitem{song_improved_score}
Y.~Song and S.~Ermon, ``Improved techniques for training score-based generative models,'' in \emph{Advances in Neural Information Processing Systems 33: Annual Conference on Neural Information Processing Systems 2020, NeurIPS 2020, December 6-12, 2020, virtual}, 2020.

\bibitem{song_sliced_2019}
Y.~Song, S.~Garg \emph{et~al.}, ``Sliced {Score} {Matching}: {A} {Scalable} {Approach} to {Density} and {Score} {Estimation},'' in \emph{Proceedings of the {Thirty}-{Fifth} {Conference} on {Uncertainty} in {Artificial} {Intelligence}, {UAI} 2019, {Tel} {Aviv}, {Israel}, {July} 22-25, 2019}, ser. Proceedings of {Machine} {Learning} {Research}, vol. 115.\hskip 1em plus 0.5em minus 0.4em\relax AUAI Press, 2019, pp. 574--584.

\bibitem{drumgan}
J.~Nistal, S.~Lattner \emph{et~al.}, ``{DRUMGAN:} synthesis of drum sounds with timbral feature conditioning using generative adversarial networks,'' in \emph{Proceedings of the 21th International Society for Music Information Retrieval Conference (ISMIR)}, Oct. 2020, pp. 590--597.

\bibitem{comparing}
------, ``Comparing representations for audio synthesis using generative adversarial networks,'' in \emph{28th European Signal Processing Conference (EUSIPCO)}.\hskip 1em plus 0.5em minus 0.4em\relax {IEEE}, Jan. 2020, pp. 161--165.

\bibitem{unet}
O.~Ronneberger, P.~Fischer \emph{et~al.}, ``U-net: Convolutional networks for biomedical image segmentation,'' in \emph{Medical Image Computing and Computer-Assisted Intervention - {MICCAI} 2015 - 18th International Conference Munich, Germany, October 5 - 9, 2015, Proceedings, Part {III}}, ser. Lecture Notes in Computer Science, vol. 9351.\hskip 1em plus 0.5em minus 0.4em\relax Springer, 2015, pp. 234--241.

\bibitem{speech_enhancement}
J.~Richter, S.~Welker \emph{et~al.}, ``Speech enhancement and dereverberation with diffusion-based generative models,'' \emph{{IEEE} {ACM} Trans. Audio Speech Lang. Process.}, vol.~31, pp. 2351--2364, 2023.

\bibitem{zhu2023edmsound}
G.~Zhu, Y.~Wen \emph{et~al.}, ``Edmsound: Spectrogram based diffusion models for efficient and high-quality audio synthesis,'' \emph{arXiv preprint arXiv:2311.08667}, 2023.

\bibitem{vae}
D.~P. Kingma and M.~Welling, ``Auto-encoding variational bayes,'' in \emph{2nd International Conference on Learning Representations ({ICLR})}, Apr. 2014.

\bibitem{transformer}
A.~Vaswani, N.~Shazeer \emph{et~al.}, ``Attention is all you need,'' in \emph{Advances in Neural Information Processing Systems 30}, Dec. 2017, pp. 5998--6008.

\bibitem{huber_loss}
P.~Charbonnier, L.~Blanc-Feraud \emph{et~al.}, ``Deterministic edge-preserving regularization in computed imaging,'' \emph{IEEE Transactions on Image Processing}, vol.~6, no.~2, pp. 298--311, 1997.

\bibitem{karras_elucidating_2022}
T.~Karras, M.~Aittala \emph{et~al.}, ``Elucidating the {Design} {Space} of {Diffusion}-{Based} {Generative} {Models},'' Oct. 2022, arXiv:2206.00364 [cs, stat].

\bibitem{easy_consistency_models}
Z.~Geng, W.~Luo \emph{et~al.}, ``Consistency models made easy,'' 2024.

\bibitem{swish}
P.~Ramachandran, B.~Zoph \emph{et~al.}, ``Searching for activation functions,'' in \emph{6th International Conference on Learning Representations, {ICLR} 2018, Vancouver, BC, Canada, April 30 - May 3, 2018, Workshop Track Proceedings}.\hskip 1em plus 0.5em minus 0.4em\relax OpenReview.net, 2018.

\bibitem{sd3}
P.~Esser, S.~Kulal \emph{et~al.}, ``Scaling rectified flow transformers for high-resolution image synthesis,'' \emph{arXiv preprint arXiv:2403.03206}, 2024.

\bibitem{radam}
L.~Liu, H.~Jiang \emph{et~al.}, ``On the variance of the adaptive learning rate and beyond,'' in \emph{8th International Conference on Learning Representations, {ICLR} 2020, Addis Ababa, Ethiopia, April 26-30, 2020}.\hskip 1em plus 0.5em minus 0.4em\relax OpenReview.net, 2020.

\bibitem{mtg_jamendo}
D.~Bogdanov, M.~Won \emph{et~al.}, ``The mtg-jamendo dataset for automatic music tagging,'' in \emph{Machine Learning for Music Discovery Workshop, International Conference on Machine Learning (ICML 2019)}, Long Beach, CA, United States, 2019.

\bibitem{dns}
H.~Dubey, V.~Gopal \emph{et~al.}, ``Icassp 2022 deep noise suppression challenge,'' in \emph{{IEEE} International Conference on Acoustics, Speech and Signal Processing, {ICASSP} 2022, Virtual and Singapore, 23-27 May 2022}.\hskip 1em plus 0.5em minus 0.4em\relax {IEEE}, 2022, pp. 9271--9275.

\bibitem{agostinelli_musiclm_2023}
A.~Agostinelli, T.~I. Denk \emph{et~al.}, ``{MusicLM}: {Generating} {Music} {From} {Text},'' Jan. 2023, arXiv:2301.11325 [cs, eess].

\bibitem{bassnet}
M.~Pasini, M.~Grachten \emph{et~al.}, ``Bass accompaniment generation via latent diffusion,'' in \emph{ICASSP 2024 - 2024 IEEE International Conference on Acoustics, Speech and Signal Processing (ICASSP)}, 2024, pp. 1166--1170.

\bibitem{sisdr}
J.~L. Roux, S.~Wisdom \emph{et~al.}, ``{SDR} - half-baked or well done?'' in \emph{{IEEE} International Conference on Acoustics, Speech and Signal Processing, {ICASSP} 2019, Brighton, United Kingdom, May 12-17, 2019}.\hskip 1em plus 0.5em minus 0.4em\relax {IEEE}, 2019, pp. 626--630.

\bibitem{visqol}
A.~Hines, J.~Skoglund \emph{et~al.}, ``Visqol: an objective speech quality model,'' \emph{{EURASIP} J. Audio Speech Music. Process.}, vol. 2015, p.~13, 2015.

\bibitem{visqolaudio}
C.~Sloan, N.~Harte \emph{et~al.}, ``Objective assessment of perceptual audio quality using visqolaudio,'' \emph{{IEEE} Trans. Broadcast.}, vol.~63, no.~4, pp. 693--705, 2017.

\bibitem{visqolv3}
M.~Chinen, F.~S.~C. Lim \emph{et~al.}, ``Visqol v3: An open source production ready objective speech and audio metric,'' in \emph{Twelfth International Conference on Quality of Multimedia Experience, QoMEX 2020, Athlone, Ireland, May 26-28, 2020}.\hskip 1em plus 0.5em minus 0.4em\relax {IEEE}, 2020, pp. 1--6.

\bibitem{fad}
K.~Kilgour, M.~Zuluaga \emph{et~al.}, ``Fr{\'{e}}chet audio distance: {A} reference-free metric for evaluating music enhancement algorithms,'' in \emph{20th Annual Conference of the International Speech Communication Association (INTERSPEECH)}, Sep. 2019, pp. 2350--2354.

\bibitem{clap}
Y.~Wu, K.~Chen \emph{et~al.}, ``Large-scale contrastive language-audio pretraining with feature fusion and keyword-to-caption augmentation,'' in \emph{{IEEE} International Conference on Acoustics, Speech and Signal Processing {ICASSP} 2023, Rhodes Island, Greece, June 4-10, 2023}.\hskip 1em plus 0.5em minus 0.4em\relax {IEEE}, 2023, pp. 1--5.

\bibitem{fad_correlation}
M.~Tailleur, J.~Lee \emph{et~al.}, ``Correlation of fr$\backslash$'echet audio distance with human perception of environmental audio is embedding dependant,'' \emph{arXiv preprint arXiv:2403.17508}, 2024.

\bibitem{repr_calm}
R.~Castellon, C.~Donahue \emph{et~al.}, ``Codified audio language modeling learns useful representations for music information retrieval,'' in \emph{Proceedings of the 22nd International Society for Music Information Retrieval Conference, {ISMIR} 2021, Online, November 7-12, 2021}, 2021, pp. 88--96.

\bibitem{repr_encodecmae}
L.~Pepino, P.~Riera \emph{et~al.}, ``Encodecmae: Leveraging neural codecs for universal audio representation learning,'' \emph{arXiv preprint arXiv:2309.07391}, 2023.

\bibitem{repr_pecmae}
P.~Alonso-Jim{\'e}nez, L.~Pepino \emph{et~al.}, ``Leveraging pre-trained autoencoders for interpretable prototype learning of music audio,'' \emph{arXiv preprint arXiv:2402.09318}, 2024.

\bibitem{repr_mert}
Y.~Li, R.~Yuan \emph{et~al.}, ``Mert: Acoustic music understanding model with large-scale self-supervised training,'' 2023.

\bibitem{magnatagatune}
D.~Wolff, S.~Stober \emph{et~al.}, ``A systematic comparison of music similarity adaptation approaches,'' in \emph{Proceedings of the 13th International Society for Music Information Retrieval Conference, {ISMIR} 2012, Mosteiro S.Bento Da Vit{\'{o}}ria, Porto, Portugal, October 8-12, 2012}.\hskip 1em plus 0.5em minus 0.4em\relax {FEUP} Edi{\c{c}}{\~{o}}es, 2012, pp. 103--108.

\bibitem{beatport}
Ángel Faraldo, ``Beatport edm key dataset,'' Jan. 2018.

\bibitem{tinysol}
C.~Emanuele, D.~Ghisi \emph{et~al.}, ``{TinySOL: an audio dataset of isolated musical notes},'' Jan. 2020.

\bibitem{musicnn}
J.~Pons and X.~Serra, ``musicnn: Pre-trained convolutional neural networks for music audio tagging,'' \emph{arXiv preprint arXiv:1909.06654}, 2019.

\bibitem{clmr}
J.~Spijkervet and J.~A. Burgoyne, ``Contrastive learning of musical representations,'' in \emph{Proceedings of the 22nd International Society for Music Information Retrieval Conference, {ISMIR} 2021, Online, November 7-12, 2021}, 2021, pp. 673--681.

\bibitem{christos_thesis}
C.~Plachouras, ``{Beyond Benchmarks: A Toolkit for Music Audio Representation Evaluation},'' Ph.D. dissertation, Universitat Pompeu Fabra, Sep. 2023.

\bibitem{mir_ref}
C.~Plachouras, P.~Alonso-Jim\'enez \emph{et~al.}, ``mir\_ref: A representation evaluation framework for music information retrieval tasks,'' in \emph{37th Conference on Neural Information Processing Systems (NeurIPS), Machine Learning for Audio Workshop}, New Orleans, LA, USA, 2023.

\end{thebibliography}

% For non bibtex users:
%\begin{thebibliography}{citations}
% \bibitem{Author:17}
% E.~Author and B.~Authour, ``The title of the conference paper,'' in {\em Proc.
% of the Int. Society for Music Information Retrieval Conf.}, (Suzhou, China),
% pp.~111--117, 2017.
%
% \bibitem{Someone:10}
% A.~Someone, B.~Someone, and C.~Someone, ``The title of the journal paper,''
%  {\em Journal of New Music Research}, vol.~A, pp.~111--222, September 2010.
%
% \bibitem{Person:20}
% O.~Person, {\em Title of the Book}.
% \newblock Montr\'{e}al, Canada: McGill-Queen's University Press, 2021.
%
% \bibitem{Person:09}
% F.~Person and S.~Person, ``Title of a chapter this book,'' in {\em A Book
% Containing Delightful Chapters} (A.~G. Editor, ed.), pp.~58--102, Tokyo,
% Japan: The Publisher, 2009.
%
%
%\end{thebibliography}

\end{document}